\documentclass[aps,prl,twocolumn,showpacs,groupedaddress]{revtex4}
\usepackage{graphicx} 
\usepackage{amsmath}
\usepackage{longtable}

\begin{document}

\title{The existence of a $^2$P$^o$ excited state for the $e^+$Ca system}
\author{M.W.J.Bromley}
  \email{mbromley@physics.sdsu.edu}
\affiliation{Department of Physics, San Diego State University, San Diego CA 92182, USA}
\author{J.Mitroy}
  \email{jxm107@rsphysse.anu.edu.au}
\affiliation{Faculty of Technology, Charles Darwin University, Darwin NT 0909, Australia}

\date{\today}

\begin{abstract}

The Configuration Interaction method is used to demonstrate that 
there is an electronically stable state of positronic calcium with 
an orbital angular momentum of $L = 1$.  This prediction relies on 
the use of an asymptotic series to estimate the variational limit of 
the energy.  The best estimate of the binding energy is $37$ meV.  A 
discussion of the structure of the system is also presented.

\end{abstract}


\maketitle

In 1997 the existence of positron-atom bound states was demonstrated 
by two independent calculations \cite{ryzhikh97b,strasburger98} of the
$e^+$Li ground state.  Subsequently, it has been shown that at least
nine other atoms can attach a positron and form an electronically stable
bound state \cite{mitroy02b}.  Besides its intrinsic interest, the
knowledge that positrons can form bound states has been crucial to 
recent developments in understanding the very large annihilation
rates that occur when positrons annihilate with various molecules 
in the gas phase \cite{iwata00a,gilbert02a,mitroy02c,marler04a}.  
The problem of explaining the large annihilation rates had
remained essentially unresolved almost since the first experiments    
\cite{paul63a,charlton85,surko88}.  While the possible 
influence of bound states upon the annihilation rate had been 
conjectured \cite{goldanskii64a,dzuba96},
the lack of hard evidence for the existence of positron-atom bound
states had certainly inhibited development of compound state models  
of positron-molecule annihilation 
\cite{dzuba96,gribakin00a,reich04a,barnes03a}.  The prevailing view on 
positron binding \cite{schrader98} has changed to such an extent over 
the last decade that a positron-atom (or positron-molecule) interaction 
potential that supports a positron bound state can now be regarded as 
mundane \cite{nishimura04b}.  

One feature of the atomic calculations is that binding has only been 
seen for spherically symmetric states.  The angular momentum 
of the parent atom ground state and the positron-atom composite state  
are always zero \cite{mitroy02b}.  Another feature is that binding 
occurs to atoms with an ionization energy close to 6.80 eV (the Ps 
binding energy) and the binding energies are largest for atoms with 
their ionization energies closest to 6.80 eV \cite{mitroy02b}.     

While the existence of positron binding to atoms (and molecules) is now 
accepted, the question of whether these complexes have excited states is 
largely unexplored.   Whether such states exist is best determined 
by calculations that are sufficiently sophisticated to accurately model 
the delicate interplay of attractive and repulsive coulomb interactions 
with the additional complication of an angular momentum barrier.  
(We note the prediction of a $^2$P$^o$ state of $e^+$Mg
by Gribakin {\em et al} \cite{gribakin96}.  However, the many body theory
of Gribakin {\em et al} is known to grossly overestimate the strength of the
positron-atom interaction \cite{mitroy01c,mitroy02b,bromley02b,bromley06c}
and so the result has never been taken seriously.)
First, it is necessary to identify what is meant by an excited state.  The states 
of interest should have the same long-range dissociation channel as the 
lower lying positronic atom ground state.  This is to distinguish these 
systems from states which are better described as a positron bound to an 
excited state of the atom (an example is the metastable $e^+$He($^3$S$^e$) 
state \cite{ryzhikh98d}).  

The present letter describes some very large configuration interaction (CI) 
calculations of the $e^+$Ca system that indicate the presence of a $^2$P$^o$ 
state with a binding energy of $\approx \! 37$ meV with respect 
to the lowest energy Ca$^+$($4s$) +  Ps($1s$) dissociation channel (the 
ionization energy of the Ca atom is less than 6.80 eV).  This system 
is the first representative of a new class of positron-atom 
bound states whose existence is more surprising than that of the 
Ca$^-$ $^2$P$^o$ negative ion \cite{pegg87,froese87}.   

The CI method as applied to positron-atom systems with two
valence electrons and a positron has been discussed previously  
\cite{bromley02a,bromley02b,mitroy06a}, but a short description 
is worthwhile.  The model Hamiltonian is initially based on a 
Hartree-Fock (HF) wave function for the neutral atom ground state.
One- and two-body semi-empirical polarization potentials are added 
to the potential field of the HF frozen-core and the parameters of 
the core-polarization potentials defined by reference to 
the spectrum of Ca$^+$ \cite{bromley02b}.
The effective Hamiltonian for the system 
with 2 valence electrons and a positron was 
\begin{eqnarray}
H  &=&  - \frac{1}{2}\nabla_{0}^2 - \sum_{i=1}^{2} \frac {1}{2} \nabla_{i}^2 
- V_{\rm dir}({\bf r}_0) + V_{\rm p1}({\bf r}_0) \nonumber \\  
&+& \sum_{i=1}^{2} \left( V_{\rm dir}({\bf r}_i) + V_{\rm exc}({\bf r}_i) + V_{\rm p1}({\bf r}_i) \right) 
    - \sum_{i=1}^{2} \frac{1}{r_{i0}} \nonumber \\  
   &+& \sum_{i<j}^{2} \frac{1}{r_{ij}}  
   - \sum_{i<j}^{2} V_{\rm p2}({\bf r}_i,{\bf r}_j)
   + \sum_{i=1}^{2} V_{\rm p2}({\bf r}_i,{\bf r}_0) \ .
\end{eqnarray}
The direct potential ($V_{\rm dir}$) represents the interaction 
with the HF $1s^22s^22p^63s^23p^6$ electron core.  The direct part of the core 
potential is attractive for electrons and repulsive for the 
positron.  The exchange potential  ($V_{\rm exc}$) between the 
valence electrons and the HF core was computed without 
approximation.

The one-body polarization potential ($V_{\rm p1}$) was a semi-empirical
polarization potential with the functional form
\begin{equation}
V_{\rm p1}(r)  =  -\sum_{\ell m} \frac{\alpha_d g_{\ell}^2(r)}{2 r^4} 
                    |\ell m \rangle \langle \ell m| .
                                    \label{polar1}
\end{equation}
The factor $\alpha_d$ is the static dipole polarizability of the 
core and $g_{\ell}^2(r) = 1-\exp\bigl(-r^6/\rho_{\ell}^6 \bigr)$ 
is a cutoff function designed to make 
the polarization potential finite at the origin.  The core dipole 
polarizability was set to 3.16 $a_0^3$ while the $\rho_{\ell}$ 
were adjusted to reproduce the Ca$^+$ spectrum \cite{bromley02b} 
(the Ca$^+$ energy is $-0.43628653$ Hartree in the model potential 
while experiment gives $-0.436278$ Hartree \cite{bashkin75a}). 
The same cutoff function has been adopted for both the positron 
and electrons.  
The two-body polarization potential ($V_{\rm p2}$) is defined as
\begin{equation}
V_{\rm p2}({\bf r}_i,{\bf r}_j) = \frac{\alpha_d} {r_i^3 r_j^3}
({\bf r}_i\cdot{\bf r}_j)g_{\rm p2}(r_i)g_{\rm p2}(r_j)\ . 
                                    \label{polar2}    
\end{equation}
where $g_{\rm p2}(r)$ is chosen to have a cutoff parameter obtained by 
averaging the $\rho_{\ell}$.  This model has been used to describe the  
calcium spectrum to quite high accuracy \cite{bromley02b,mitroy03f}.

The CI basis was constructed by 
letting the two electrons and the positron form all the possible
total angular momentum $L_T = 1$ configurations, with the
two electrons in a spin-singlet state, subject to the selection rules,
\begin{eqnarray}
\max(\ell_0,\ell_1,\ell_2) & \le & J \ , \\
\min(\ell_1,\ell_2)& \le & L_{\rm int} \ ,  \\  
(-1)^{(\ell_0+\ell_1+\ell_2)}& = & -1  \ . 
\end{eqnarray}
In these rules $\ell_0$, $\ell_1$ and $\ell_2$ are respectively 
the orbital angular momenta of the positron and the two electrons.  
We define $\langle E \rangle_J$ to be the energy of the calculation 
with a maximum orbital angular momentum of $J$.  

The two-electron-positron calculations with non-zero total angular 
momentum were first validated against the previous $L_T =1$ 
PsH calculations of Tachikawa \cite{tachikawa01a}.  Using their 
Gaussian-type orbitals we reproduced their reported energy and
annihilation rates (note that the PsH states with $L_T > 0$ are
unbound).

\begin{table}[th]
\caption[]{  \label{tab:calmax}
Results of CI calculations for the $e^+$Ca $^2$P$^o$ state as 
a function of $J$, and for $L_{\rm int}=3$.  The total number 
of configurations  is denoted by $N_{CI}$.  The 3-body energy 
of the state, relative to the energy of the Ca$^{2+}$ core, 
is given in Hartree.  The threshold for binding is $-0.68628653$ 
Hartree, and $\varepsilon$ gives the binding energy (in Hartree) 
against dissociation into Ps + Ca$^+$($4s$).  The values of 
$\langle E \rangle_{\infty}$ were determined at $J = 14$.
}
\begin{ruledtabular}
\begin{tabular}{lccc} 
$J$&   $N_{CI}$ & $\langle E \rangle_{J}$ & $\varepsilon_J$  \\ \hline
  1 &   10094 & -0.64319380 & -0.04309273   \\
  2 &   34244 & -0.64976077 & -0.03652576   \\
  3 &   79198 & -0.65519252 & -0.03109401   \\
  4 &  140168 & -0.66104645 & -0.02524009   \\
  5 &  206822 & -0.66626528 & -0.02002126  \\
  6 &  278754 & -0.67046020 & -0.01582633   \\
  7 &  352156 & -0.67372823 & -0.01255830   \\
  8 &  426832 & -0.67626613 & -0.01002041   \\
  9 &  501508 & -0.67824946 & -0.00803707   \\
 10 &  576184 & -0.67981518 & -0.00647135   \\
 11 &  650860 & -0.68106444 & -0.00522209   \\
 12 &  725536 & -0.68207134 & -0.00421520   \\
 13 &  800212 & -0.68289035 & -0.00339618   \\
 14 &  874888 & -0.68356185 & -0.00272468   \\
\hline  
 &   &   \multicolumn{1}{c}{$\langle E \rangle_{\infty}$} &  \multicolumn{1}{c}{$\varepsilon_{\infty}$}   \\
\multicolumn{2}{l}{1-term eq.(\ref{extrap1})}   & -0.68648706 &  0.0002005  \\
\multicolumn{2}{l}{2-term eq.(\ref{extrap1})}   & -0.68739784 &  0.0011113  \\
\multicolumn{2}{l}{3-term eq.(\ref{extrap1})}   & -0.68763826 &  0.0013517  \\
\end{tabular}
\end{ruledtabular}
\end{table} 

The Hamiltonian for the $e^+$Ca $^2$P$^o$ state was
diagonalized in a CI basis constructed from a very large number
of single particle orbitals, including orbitals up to $\ell = 14$.
There was a minimum of $14$ radial basis functions for each $\ell$.
The largest calculation was performed with $J = 14$ and
$L_{\rm int} = 3$ and gave a CI basis dimension of 874888.  
The parameter $L_{\rm int}$ does not have to be particularly large
since it is mainly concerned with electron-electron correlations 
\cite{bromley02b}.  The resulting Hamiltonian matrix was diagonalized 
with the Davidson algorithm \cite{stathopolous94a}, and a total 
of 4000 iterations were required for the largest calculation.   

\begin{figure}[tbh]
\centering{
\includegraphics[width=9.0cm,angle=0]{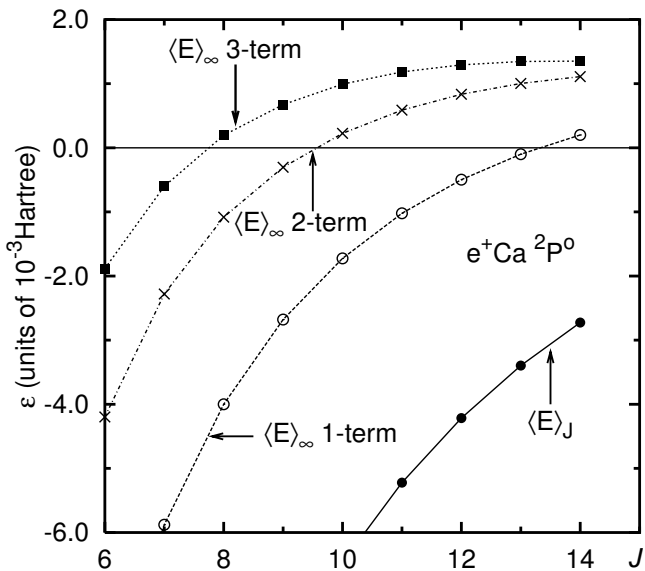}
}
\caption[]{ \label{fig:CaE}
The binding energy (in units of Hartree) of the $^2$P$^o$ state 
of $e^+$Ca as a function of $J$.  The directly 
calculated energy is shown as the solid line while the 
$J \to \infty$ limits using eq.~(\ref{extrap1}) 
with 1, 2 or 3 terms are shown as the dashed lines.     
The Ca$^+$($4s$) + Ps($1s$) dissociation threshold is shown 
as the horizontal solid line.
}
\end{figure}

The energy of the $e^+$Ca $^2$P$^o$ state as a function of $J$ is
given in Table \ref{tab:calmax}.  The binding energy is defined as
$\varepsilon = -(0.68628653 + E)$.  None of the calculations give an
energy lower than the Ca$^+$($4s$) + Ps($1s$) threshold.  The main technical 
problem afflicting CI calculations of positron-atom interactions is the 
slow convergence of the energy with $J$ 
\cite{mitroy99c,dzuba99,mitroy02b,mitroy06a} and the present calculation 
is no exception to the rule.  One way to 
determine the $J \rightarrow \infty$ energy, $\langle E \rangle_{\infty}$,
is to make use of an asymptotic analysis.  It has been shown that successive 
increments, $\Delta E_{J} = \langle E \rangle_J - \langle E \rangle_{J-1}$, 
to the energy can written as an inverse power series 
\cite{schwartz62a,carroll79a,hill85a,mitroy06a,bromley06a}, viz 
\begin{equation}
\Delta E_J \approx \frac {A_E}{(J+{\scriptstyle \frac{1}{2}})^4} 
    + \frac {B_E}{(J+{\scriptstyle \frac{1}{2}})^5} 
    + \frac {C_E}{(J+{\scriptstyle \frac{1}{2}})^6} + \dots \ \   .
\label{extrap1}
\end{equation}
The $J \to \infty$ limit, has been determined by fitting sets of 
$\langle E \rangle_J$ values to asymptotic series with either 1, 2 
or 3 terms.  The coefficients, $A_E$, $B_E$ and $C_E$ for the 3-term 
expansion are determined at a particular $J$ from 4 successive energies
($\langle E \rangle_{J-3}$, $\langle E\rangle_{J-2}$, 
$\langle E \rangle_{J-1}$ and $\langle E \rangle_{J}$).  
Once the coefficients have been determined it
is easy to sum the series to $\infty$ and obtain the 
variational limit.   Application of asymptotic series analysis
to helium has resulted in CI calculations reproducing the ground
state energy to an accuracy of $\approx \!\! 10^{-8}$ Hartree
\cite{salomonson89b,bromley06a}.  Figure \ref{fig:CaE} shows the 
estimates of $\langle E \rangle_{\infty}$ as a function of $J$.    

The different extrapolations all give energies below the dissociation 
threshold and indicate that the $e^+$Ca $^2$P$^o$ state 
is electronically stable.  The energy of the 
three-term extrapolation does seem to have stabilized at a
binding energy of $\approx \!\! 0.00135$ Hartree ($37$ meV).  The two-term binding
energy is slightly smaller but does seem to be approaching the 
three-term estimate.  The one-term estimate of $\langle E \rangle_{\infty}$  
is also bound, although its binding energy is 
smaller.  The precise estimates
of $\langle E \rangle_{\infty}$ evaluated at $J = 14$  
are given in Table \ref{tab:calmax}.  

Since the binding energy is small it is desirable to examine
the areas of uncertainty in the model and computation to determine
whether they could invalidate the prediction.      

The interaction between the core and valence electrons was tested
quite simply by adjusting the core polarizability by $\pm5\%$ (leading
to a change of $\pm 0.16\%$ in the neutral Ca ionization energy).
When this was done, the overall change in the $e^+$Ca $^2$P$^o$ 
binding energy at $J = 14$ was $\pm 0.00013$ Hartree, i.e.
$\pm 10\%$ of the final binding energy.      

Choosing the polarization potential cutoff function for the positron 
to be the same as the electron will lead to the binding energy 
being underestimated.  First, it is known from calculations on the 
rare gases that the positron polarization potential is more 
attractive than the equivalent electron potential 
\cite{mitroy02d,mitroy03b,mitroy03c}.  Also, the \textit{ab-initio} 
calculations on the small systems, $e^+$He($^3$S$^e$) 
and $e^+$Li have given larger binding energies (by 1-2$\%$) than 
calculations using model potentials to represent the core 
\cite{mitroy04c,mitroy05e}. 
  
The lack of completeness in the finite dimension radial basis is 
also not an issue.  Computational investigations have revealed that
accurate prediction of the $\Delta E_J$ energy increments 
requires a larger basis as $J$ increases \cite{bromley06a,mitroy06a}.  
This results in the typical CI partial wave expansion with a fixed 
dimension radial basis for the different $\ell$-values having an inherent 
tendency to underestimate the binding energy \cite{bromley06a,mitroy06a}.  

\begin{figure}[bht]
\centering{
\includegraphics[width=9.0cm,angle=0]{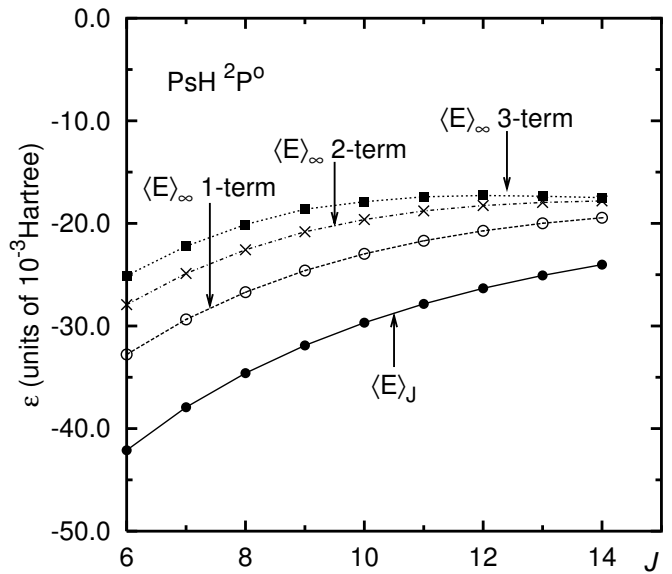}
}
\caption[]{ \label{fig:PsHE}
The binding energy, $\varepsilon = -(0.75 + E)$, of the PsH $^2$P$^o$ 
system as a function of $J$.  The directly calculated energy is shown 
as the solid line while the $\langle E \rangle_{\infty}$ 
limits using eq.~(\ref{extrap1}) are shown as the dashed 
lines.   
}
\end{figure}

Finally, a computational null experiment was performed on the PsH system.  
This system does not have a $^2$P$^o$ bound state.  A calculation of 
almost identical size to the $e^+$Ca system was performed.  An unbound 
system would be expected to have an $\langle E \rangle_\infty$ that 
asymptotes to the threshold energy, or to a value above threshold, and 
this is what is seen to occur in Figure \ref{fig:PsHE}.  

A $e^+$Ca valence annihilation rate of $\Gamma = 1.42 \times 10^9$ 
sec$^{-1}$ has also been determined using an asymptotic analysis 
similar to that used for the energy \cite{bromley06c,mitroy06a}.  
This large annihilation rate suggests that a large fraction of the   
wave function consists of a Ca$^+ +$Ps($1s$) cluster \cite{mitroy02b}.  

The system is compact despite its small binding energy and the 
mean positron radius for a converged calculation was estimated 
at  $\langle r_p \rangle \approx 8.7$ $a_0$.   
The $e^+$Ca ground state with a binding energy 14 times
larger has almost the same $\langle r_p \rangle$ \cite{bromley06c}. 
However, the large $r$ form of the  $^2$P$^o$ wave function must have a  
Ca$^+$($4s$) + Ps($1s$) structure with the Ps($1s$) center of mass being 
in an $L = 1$ state with respect to the residual ion.  The centrifugal
barrier associated with the non-zero angular momentum acts to
confine the positron probability distribution.        

The present calculations indicate that positronic calcium has a 
$^2$P$^o$ excited state.  The existence of both $^2$S$^e$ and 
$^2$P$^o$ states of $e^+$Ca makes optical detection a
possibility.  While the present calculation does not present 
an absolute variational proof of binding (the calculation 
would have to be extended to $J \approx 20$ for this to occur), 
the evidence in support of the excited state is strong.  It is 
worth noting that extrapolating finite dimension basis sets to 
the variational limit is quite common in the field of quantum 
chemistry \cite{klopper99a}.  

One consequence of this result lies in the area of positron 
annihilation.  It has been shown that a low energy $p$-wave   
shape resonance can lead to very large values of $Z_{\rm eff}$   
\cite{bromley03a}.  It is possible for thermally averaged values 
of $Z_{\rm eff}$ to exceed 10$^4$ since the energy dependence 
of $Z_{\rm eff}$ for a $p$-wave shape resonance is reasonably 
compatible with a Maxwell-Boltzmann energy 
distribution.  The existence of $p$-wave shape resonances are 
certainly plausible given the existence of the $^2$P$^o$ bound 
state and provides another reaction that can contribute to 
the very large annihilation rates seen in gas-phase 
experiments \cite{paul63a,charlton85,surko88,iwata00a}.  
And very recently, $Z_{\rm eff}$ peaks in the annihilation spectra for
dodecane (C$_{12}$H$_{26}$) and tetradecane (C$_{14}$H$_{30}$)  
have been tentatively identified as a positronically excited
bound state associated with the C-H stretch mode 
\cite{barnes06a,gribakin06a}. 
 
The authors would like to thank Dr. Masanori Tachikawa for
access to unpublished data on systems with non-zero angular momentum.
These calculations were performed on Linux clusters hosted at the 
South Australian Partnership for Advanced Computing (SAPAC) and 
SDSU Computational Sciences Research Center, with technical
advice given by Grant Ward, Patrick Fitzhenry and Dr James Otto.  

% \bibliography{positron}

\end{document}